\documentclass[acus]{JAC2000}

\usepackage{graphicx}
\usepackage{citesort}       
\usepackage[]{amsmath}
\usepackage[]{amssymb}

\begin{document}
\title{ELECTRO-OPTIC SAMPLING OF TRANSIENT FIELDS FROM THE PASSAGE OF HIGH-CHARGE 
ELECTRON BUNCHES\thanks{Work supported in part by Fermilab 
which is operated 
by URA, Inc. for the U.S. DoE under contract
\mbox{DE-AC02-76CH03000}. Current address of P. L. Colestock is LANL. 
e-mail:~\mbox{mjfitch@pas.rochester.edu}}}

\author{M. J. Fitch, A. C. Melissinos, University of Rochester, 
Rochester NY 14627, USA\\
P. L. Colestock, 
J.-P. Carneiro, H. T. Edwards, W. H. Hartung, FNAL, 
Batavia IL 60510, USA}

\maketitle

\begin{abstract} 
When a relativistic electron bunch traverses a structure, strong electric 
fields are induced in its wake. We present measurements of 
the electric field as a function of time as measured at a fixed location 
in the beam line.  For a 12~nC bunch of duration 4.2~ps FWHM, the peak 
field is measured $>0.5$~MV/m. Time resolution of $\sim$5~ps is achieved using 
electro-optic sampling with a lithium tantalate (LiTaO$_{3}$) crystal and a short-pulse 
infrared laser synchronized to the beam. We present measurements under 
several different experimental conditions and discuss the influence of 
mode excitation in the structure.
\end{abstract}

\section{INTRODUCTION}

Since the pioneering experiments \cite{Valdmanis:1982,Valdmanis:1983,Auston:1988},
 electro-optic sampling (EOS) has been shown to be
a powerful technique for fast time-domain measurements of electric 
fields \cite{Wu:1997b,Leitens:1999}.

The use of electro-optic sampling for accelerator applications has 
been previously suggested by \cite{Channell:1982,Pavlov:1982,Geitz:1997} 
and others. 
Detection of the beam current by 
magneto-optic effects has been demonstrated by \cite{Pavlov:1982} with
a time resolution that is subnanosecond.

Recently, at Brookhaven, electro-optic 
detection of a charged particle beam was reported by detecting a 
faint light pulse through crossed polarizers as the beam passed by an 
electro-optic crystal \cite{Semertzidis-x:1999}. The time resolution possible here
is limited by the speed of the photodetectors and amplifiers, which 
similar to that available with capacitive beam pickups ($\sim$100~ps). Earlier at 
Brookhaven, an RF phase measurement using the electro-optic effect 
and phase stabilization by feedback was demonstrated \cite{Leung:1993}.

We have used electro-optic 
sampling to measure the electric field waveforms in vacuum induced by 
the passage of electron bunches with an estimated time resolution of 
$\sim$5~ps, limited by the laser pulse length 
\cite{Fitch-UR-x:1999,Fitch-PAC-x:1999}. 

Independently of our work, a group at FOM 
Rijnhuizen (Nieuwegein, The Netherlands) has used 
electro-optic sampling in ZnTe to resolve the sinusoidal electric field of 
the free electron laser FELIX at the 
optical frequency ($\lambda=150~\mu$m) \cite{Knippels-x:1999}. Of note 
is the rapid-scanning cross-correlation technique (a fast 
data-acquisition trick). The same group has 
sampled the electric field of the transition radiation from the 
electron beam exiting a beryllium window \cite{Oepts-x:1999} and the 
electric field in 
vacuum \cite{Yan-x:2000} from which the bunch length is measured.
 
We have thus far been unable to reproduce their results with ZnTe; 
we suspect a problem with our crystal.

\section{EXPERIMENT}

The linear electro-optic effect (or Pockels effect) is one of several 
nonlinear optical effects that arise from the second-order 
susceptibility tensor $\mathbf{\chi}^{(2)}$, and is described in many standard 
texts, such as \cite{Yariv:1985}. For our purposes, it suffices that 
the polarization of light is altered by an electric field applied to 
the crystal. By analyzing the polarization change, the electric field 
can be measured. Using a short laser pulse and a thin crystal, the 
electric field is sampled at a particular time T$_{i}$ when the laser pulse 
arrives at the crystal.
By changing the delay of the probe laser 
arrival time, and repeatedly measuring the electric field, the 
electric field waveform is recovered by electro-optic sampling. The 
data acquistion is handled by LabVIEW and a digital oscilloscope.

\begin{figure}[htb]
\centering
\includegraphics*[width=65mm,bbllx=14mm,%
    bburx=185mm,bblly=52mm,bbury=256mm]{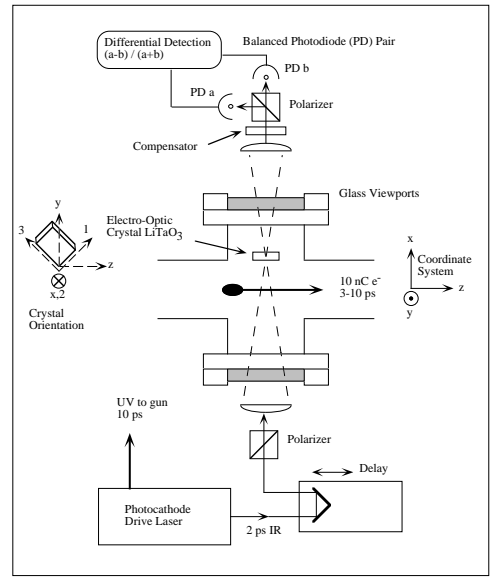}
\caption{EOS configuration, sensitive to $(E_{z}+E_{\theta})/\sqrt{2}$.}
\label{fig:MOC05-5}
\end{figure}

Experiments were performed 
at the A{\O} Photoinjector of Fermilab \cite{Colby:1997,Carneiro-x:1999}. 
A lamp-pumped Nd:glass laser 
system built by the University of Rochester is quadrupled to UV 
($\lambda=263$~nm) for photocathode excitation. The UV pulses are 
temporally shaped to an approximate flat-top distribution with a 
10.7~ps FWHM. Unconverted infrared 
light is the probe laser for the electro-optic sampling, so that 
jitter between the beam and the probe laser vanishes to first order. 
The photoinjector produces 12~nC bunches with normalized emittance of 
20$\pi$~mm-mrad (uncompressed) in pulse trains up to 200 pulses long with interpulse 
spacing 1~$\mu$sec. A chicane of four dipoles was used for magnetic 
compression. In a companion paper in these proceedings we present some 
compression studies. The best compression to date is 
$\sigma_{z}=0.63$~mm (1.89~ps) for a charge of
13.2~nC, which gives a peak current of 2.8~kA.

\begin{figure}[htb]
\centering
\includegraphics*[width=65mm]{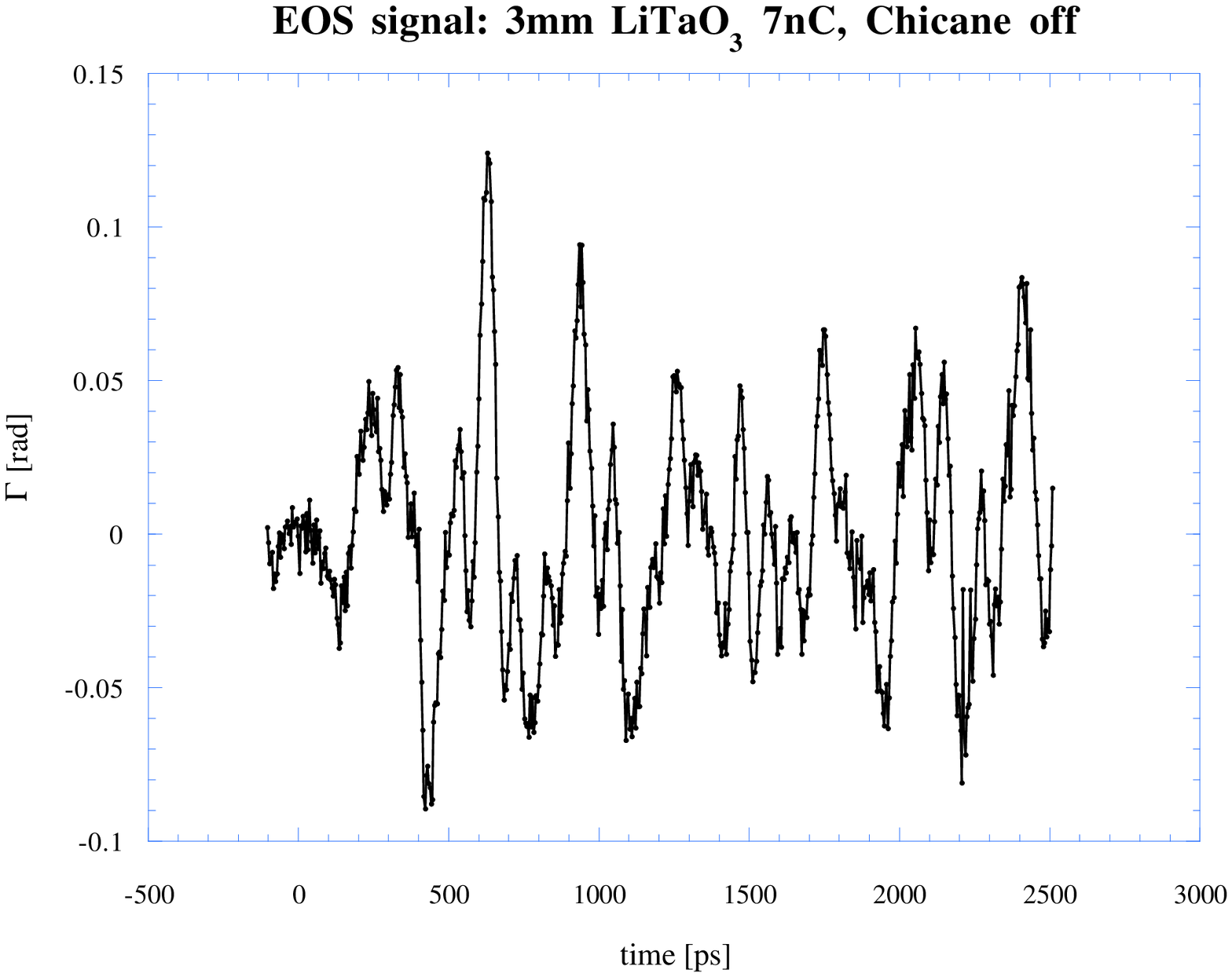}
\caption{EOS waveform, sensitive to $(E_{z}+E_{\theta})/\sqrt{2}$.}
\label{fig:MOC05-1}
\end{figure}

\begin{figure}[htb]
\centering
\includegraphics*[width=65mm]{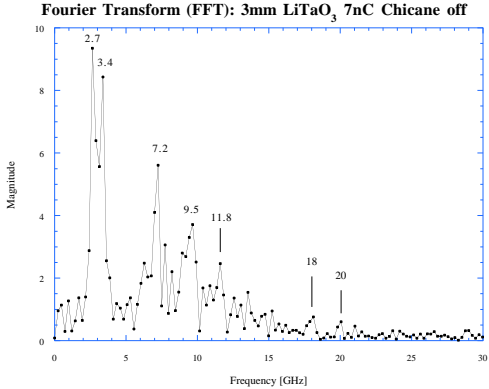}
\caption{FFT of waveform in Figure \ref{fig:MOC05-1}}
\label{fig:MOC05-2}
\end{figure}

We have taken data using several different configurations.
The elements common to all of the setups are a polarizer, the 
crystal, the compensator, and another polarizer (analyzer).
The ellipsometry can be simplified for perfect polarizers and small
polarization changes in the crystal. For two detectors $A$ and $B$ 
(silicon photodiodes) after the analyzer, the intensity measured at $I_{A}\equiv A$ is:
\begin{equation}
    A= I_{o} \sin^{2}(\delta\Gamma + \phi)
    \label{eq:A}
\end{equation}
where the intensity incident on the analyzer is $I_{o}$, and $\phi$ 
is a constant which represents the compensator and/or the static 
birefringence of the crystal 
($\phi_{s}=\omega(n_{o}-n_{e})L/c$). The term proportional to the electric 
field is $\delta\Gamma = \omega\delta n \, L/c$, and 
putting in the electro-optic coefficient for LiTaO$_{3}$ with the 
electric field along the 3-axis, we find
\begin{equation}
    \delta\Gamma = \frac{\omega}{c} 
    (n_{o}^{3}\,r_{13}-n_{e}^{3}\,r_{33}) E_{3} L.
    \label{eq:dGamma3}
\end{equation}
For the electric field along the 2-axis of LiTaO$_{3}$, the electro-optic 
coefficient is $\delta\Gamma =\omega n_{o}^{3}\,r_{22}\, E_{2} L/c$.
It is clear from Equation \ref{eq:A} that  if $\phi=0$, then for 
small signals, $A \propto I_{o} (\delta\Gamma)^{2}$. 

The second detector $B$ measures the orthogonal polarization 
component, so $ B= I_{o} \cos^{2}(\delta\Gamma + \phi)$.
It is seen that for a choice of $\phi=\pi/4$,
\begin{equation}
    \frac{A-B}{A+B} = \sin\delta\Gamma \sim\delta\Gamma \propto E
    \label{eq:diff-over-sum}
\end{equation}
independent of $I_{o}$. The compensator then is used to balance the 
detectors in the absence of electro-optic modulation. However, the 
static birefringence is a function of temperature, so we make one 
further subtraction to cancel drifts to form the experimental $\Gamma$.
\begin{equation}
    \Gamma = \left(\frac{A-B}{A+B}\right)_{\mathrm{signal}} \!\!-\,\,
             \left(\frac{A-B}{A+B}\right)_{\mathrm{background}}
    \label{eq:exp-Gamma}
\end{equation}
For the background points, a shutter is closed which blocks the UV for 
the photocathode but allows the infrared probe laser to go to the 
crystal. The field 
magnitude is estimated by calibrations on a duplicate crystal on the 
bench. A field $E_{3}=100$~kV/m induces a rotation 
$\Gamma=0.046$~rad, while $E_{2}=100$~kV/m induces $\Gamma=0.003$~rad, all for 
the 7$\times$8$\times$1.5~mm LiTaO$_{3}$ crystal (thickness $L=1.5$~mm).

\section{RESULTS}

With 
the sensitive axis of the crystal oriented so that 
$E_{3}=(E_{z}+E_{\theta})/\sqrt{2}$, using the convention 
that the electron beam velocity defines the $+z$ direction, the 
measured waveform in shown in Figure \ref{fig:MOC05-1}.
The initially surprising feature is the presence of strong 
oscillations that persist beyond the end of the delay stage (3~ns). 
These are attributed to  excitation of modes in the structure, and an 
FFT of the waveform is shown in Figure \ref{fig:MOC05-2}.
We can, for example, identify the frequencies near 3~GHz as trapped 
modes in the 6-way cross \cite{Tang:1997}.

\begin{figure}[htb]
\centering
\includegraphics*[width=65mm]{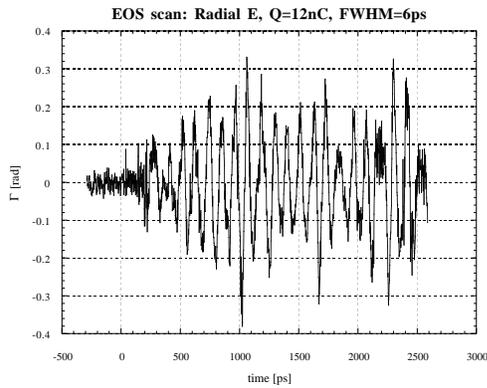}
\caption{Electro-optic sampling waveform, sensitive to $E_{r}$.}
\label{fig:MOC05-3}
\end{figure}
\begin{figure}[htb]
\centering
\includegraphics*[width=65mm]{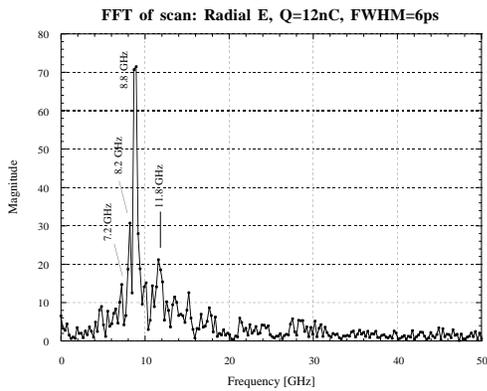}
\caption{FFT of waveform in Figure \ref{fig:MOC05-3}}
\label{fig:MOC05-4}
\end{figure}

With the sensitive axis of the crystal oriented so that $E_{3}=E_{r}$, 
the measured waveform is quite different, being nearly sinusoidal 
(Figure \ref{fig:MOC05-3}). 
In the cylindrical beam pipe (radius $b=2.2$~cm), there is a 
propagating (waveguide) TM$_{1,1}$ mode
with frequency 
$\nu = (3.83) c/2\pi b=8.4$~GHz, and it may be the origin of the 
observed 8.8~Ghz component.
The slow build-up (and beat near 1900~ps) in the 
envelope could be explained by a small splitting of this mode into two 
frequencies, which are initially out of phase. The FFT (figure 
\ref{fig:MOC05-4})
suggests a splitting, but the resolution (limited by the length of the 
scan) is poor. More will be presented and discussed in a future 
publication. A second round of experiments is planned with the goal of 
detecting the direct Coulomb field of the bunch.


The direct Coulomb field of the bunch, if detected, is simply 
connected with the charge distribution $\rho(z)$ with sensitivity to 
head-tail asymmetries. As the electro-optic effect has a physical 
response at the femtosecond level, the technique of electro-optic 
sampling could be a valuable method for bunch length measurements at 
the $<100$~fs level. The transient (wake) fields we measured off-axis 
could be applied to on-axis measurements of the wake function and beam 
impedance. Higher-order mode coupling and damping in structures may 
also be of interest.

\bibliographystyle{unsrt}
\bibliography{MOC05}

\end{document}